\documentclass[aps,floatfix,twocolumn,showpacs]{revtex4}

\usepackage[dvips]{graphicx}
\usepackage{amsfonts}
\usepackage{amssymb}

\newcommand{\ave}[1]{{\langle #1\rangle}}

\begin{document}

\title{Classical and Quantum Chaos and Control of Heat Flow}
\author{Giulio Casati${}^{1,2,3}$}
\author{Carlos Mej\'{\i}a-Monasterio${}^{4}$}
\affiliation{${}^1$Center for  Nonlinear and  Complex  Systems, Universit\`a
 degli Studi dell'Insubria, Como, Italy}
\affiliation{${}^2$CNR-INFM and Istituto Nazionale di Fisica Nucleare,
Sezione di Milano}
\affiliation{${}^3$Department of Physics, National University of Singapore,
Singapore 117542, Republic of Singapore}
\affiliation{${}^4$Dipartimento di Matematica, Politecnico di Torino,
Corso Duca degli Abruzzi 24 I-10129 Torino, Italy}

\date{\today}
\begin{abstract}
  We  discuss the  problem of  heat conduction  in classical  and  quantum low
  dimensional  systems from  a microscopic  point of  view.  At  the classical
  level we provide  convincing numerical evidence for the  validity of Fourier
  law  of heat conduction  in linear  mixing systems,  {\em i.e.}   in systems
  without  exponential instability.   At the  quantum level,  where  motion is
  characterized by the lack of exponential dynamical instability, we show that
  the validity of Fourier law is  in direct relation with the onset of quantum
  chaos. We  then study  the phenomenon of  thermal rectification  and briefly
  discuss  the different  types of  microscopic  mechanisms that  lead to  the
  rectification of heat flow.  The  control of heat conduction by nonlinearity
  opens the possibility to propose new devices such as a thermal rectifier.
\end{abstract}

\maketitle

\section{Introduction}
\label{sec:intro}

The  origin  of the  macroscopic  phenomenological  ``laws'' of  thermodynamic
transport is  still one  of the major  challenges to theoretical  physics.  In
particular the  issue of energy  (heat) transport, in  spite of having  a long
history,  is  not   completely  settled  \cite{bonetto,LLP-review}.   Given  a
particular classical,  many-body Hamiltonian system,  neither phenomenological
nor  fundamental transport  theory can  predict whether  or not  this specific
Hamiltonian  system yields  an energy  transport governed  by the  Fourier law
$J=-\kappa\nabla  T$, relating the  macroscopic heat  flux to  the temperature
gradient $\nabla T$ \cite{peierls}.  Heat flow is universally presumed to obey
a simple diffusion equation which can  be regarded as the continuum limit of a
discrete  random walk.   In consequence,  transport theory  requires  that the
underlying deterministic dynamics yield a truly random process.  Therefore, it
is not mere idle curiosity to  wonder what class, if any, of many-body systems
satisfy the necessary stringent requirements.

In     spite     of     intense     investigations    in     recent     years,
\cite{lebowitz,ford,Prosen,Kab93,leprifpu,LLP,HLZ,hatano99,Dhar01,casati85,Gendelman,alonso,Aoki01,Grassberger,triangle1,casati,mejia-1,mejia-2,emmz}
the precise conditions that a  dynamical system of interacting particles in 1D
must satisfy in order to obey the Fourier law of heat conduction are still not
known. However, the general picture which emerges is that, for systems with no
globally  conserved  quantities (i.e.,  globally  ergodic), positive  Lyapunov
exponent is  a sufficient condition to  ensure Fourier heat  law. This appears
quite natural.

Indeed  modern ergodic  theory  tells us  that  for K-systems,  a sequence  of
measurements  with  finite  precision  mimics  a  truly  random  sequence  and
therefore  these  systems  appear  precisely  those  deterministically  random
systems  tacitly   required  by  transport   theory.   In  fact   the  thermal
conductivity has been studied for  a Lorentz channel --a quasi one dimensional
billiard  with circular  scatterers-- and  it was  shown to  obey  Fourier law
\cite{alonso}.  Yet  we do not have  rigorous results and in  spite of several
efforts,  the  connection  between  Lyapunov  exponents,  correlations  decay,
diffusive  and  transport  properties  is  still not  completely  clear.   For
example, in a  recent paper \cite{triangle1} a model  has been presented which
has zero  Lyapunov exponent and  yet it exhibits unbounded  Gaussian diffusive
behavior.  Since  diffusive behavior is at  the root of  normal heat transport
then  the above  result\cite{triangle1} constitutes  a strong  suggestion that
normal heat conduction  can take place even without  the strong requirement of
exponential instability.  Moreover,  the models in \cite{alonso,triangle1} are
noninteracting and thus,  the condition of Local Thermal  Equilibrium (LTE) is
not satisfied.  Noninteracting particle  systems are certainly less realistic. 
They are  good models for gases  in the Knudsen  limit in which the  mean free
path is larger  than the characteristic length of the  container and their use
in other physical situations  is open to criticism \cite{rondoni}.  Therefore,
it  is interesting  to investigate  up  to what  extend one  can simplify  the
microscopic dynamics, and yet obtain a normal transport behavior.

An  additional  interesting  problem,  almost completely  unexplored,  is  the
derivation of Fourier law from  quantum dynamics. Indeed at the quantum level,
investigations    have    been    mainly    focused   on    linear    response
theory\cite{spin-chains}.  The  possibility  to  derive  the  Fourier  law  by
directly establishing the dependence of  $J$ on $\nabla T$ in a nonequilibrium
steady state  calls directly in question  the issue of quantum  chaos. In this
connection we  recall that  a main feature  of quantum  motion is the  lack of
exponential dynamical  instability\cite{casati86}, a property which  is at the
heart of classical dynamical chaos.

The aim of this paper is to discuss the transport of heat in several classical
and quantum microscopic  models by direct numerical simulation  of energy flow
in  the system  in contact  with thermal  baths.  At  the classical  level our
results in  section \ref{sec:fourier} provide convincing  evidence that linear
mixing systems (i.e.   with zero Lyapunov exponent) obey  Fourier law.  At the
quantum  level we study  in section  \ref{sec:quantum} an  Ising chain  of $L$
coupled spins  $1/2$ in  a tilted magnetic  field. We provide  clear numerical
evidence of  the validity of  Fourier law at  the onset of quantum  chaos.  In
section \ref{sec:rectifier} we then investigate the possibility to control the
energy transport and discuss  different microscopic mechanical models in which
thermal  rectification  can be  observed.   The  possibility  to control  heat
conduction by nonlinearity opens the way to design a thermal rectifier, i.e. a
lattice that carries heat preferentially  in one direction while it behaves as
an insulator in the opposite direction.

\section{Dynamical Instability and Fourier Law}
\label{sec:fourier}

A  Lorentz  gas  consists  of  noninteracting  point  particles  that  collide
elastically with a  set of circular scatterers on the  plane. Motivated by the
ergodicity  and mixing  properties  of  the Lorentz  gas,  in \cite{alonso}  a
channel geometry  of this model  was considered to  study the problem  of heat
conduction. By imposing  and external thermal gradient it was  found that in a
Lorentz channel the Fourier law is satisfied. Yet this system is not described
by LTE.  A modification of  this model in  which particles and  scatterers can
exchange  energy through  their  collisions appeared  in \cite{mejia-1}.  This
effective interaction  leads to the  establishment of LTE.  As  a consequence,
this model has proven to reproduce realistically macroscopic transport in many
different situations  \cite{mejia-2}.  In particular, the  validity of Fourier
law  was  verified,  indicating  that  the  interaction  among  the  different
particles is not fundamental for the observation of normal transport.

To clarify the  role of dynamical instability on the  validity of Fourier law,
let  us  consider a  two  dimensional billiard  model  which  consists of  two
parallel  lines of  length $L$  at  distance $d$  and a  series of  triangular
scatterers (Fig.  1).  In this geometry,  no particle can move between the two
reservoirs without suffering elastic collisions with the triangles.  Therefore
this model  is analogous  to the one  studied in \cite{alonso}  with triangles
instead of discs and the essential  difference is that in the triangular model
discussed here the dynamical instability  is linear and therefore the Lyapunov
exponent  is  zero.   Strong   numerical  evidence  has  been  recently  given
\cite{casati} that  the motion inside  a triangular billiard, with  all angles
irrational  with  $\pi$   is  mixing,  without  any  time   scale,  (see  also
\cite{poly}).  Moreover, an  area  preserving  map, which  was  derived as  an
approximation of the boundary map for the irrational triangle, when considered
on  the cylinder  shows a  nice Gaussian  diffusive behavior  even  though the
Lyapunov  exponent  of the  map  is  zero  \cite{triangle1}. It  is  therefore
reasonable to expect  that the motion inside the  irrational polygonal area of
Fig.  1 is diffusive thus leading to normal conductivity.

\begin{figure}[!t]
\includegraphics[width=3.2in]{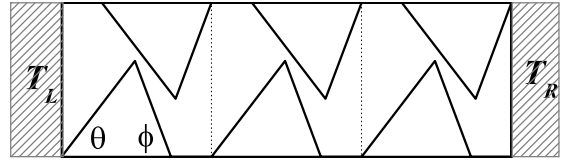}
\caption {
  The  geometry  of  the model.  Particles  move  in  the region  outside  the
  triangular scatterers.   The {\it x}  coordinate goes along the  channel and
  {\it y  } is perpendicular to  it.  The two heat  reservoirs at temperatures
  $T_L$ and $T_R$ are  indicated. The length of each cell is  $l= 3$, the base
  of the  triangles is  $a= 2.19$  and the distance  between the  two parallel
  lines is $d=1.8$.  The geometry  is then uniquely specified by assigning the
  internal angles $\theta$ and $\phi$ } \label{model}
\end{figure}

We consider a channel of total length $L =Nl$ where $N$ and $l$ are the number
and the  length of the  fundamental cells. For  the irrational angles  we take
$\theta=(\sqrt{2}-1)\pi/2$  and $\phi=1$.   By increasing  $L$, the  number of
particles per cell must be kept  constant. However, since the particles do not
interact one may consider the motion  of a single particle over long times and
then rescale the flux.

Heat conductivity of this model has been numerically studied in \cite{triang}.
Heat  baths have  been simulated  with  stochastic kernels  of Gaussian  type,
namely, the  probability distribution of  velocities for particles  coming out
from the baths is
\begin{equation} \label{eq:baths}
P(v_x) =
\frac{|v_x|}{T}\exp\left(-\frac{v^2_x}{2T}\right), \
P(v_y) = \frac{1}{\sqrt{2\pi T}}\exp\left(-\frac{v^2_y}{2T}\right)
\label{Gaussian}
\end{equation}
for $v_x$ and $v_y$, respectively.

Since the energy changes only at collisions with the heat baths, the heat flux
is given by
\begin{equation}
j (t_c)=\frac{1}{t_c}\sum_{k=1}^{N_c}(\Delta E)_k , \label{flux}
\end{equation}
where $(\Delta E)_k = E_{in} - E_{out}$  is the change of energy at the $k$-th
collision with the heat bath and  $N_c$ is the total number of such collisions
which occur  during time  $t_c$.  We have  checked that for  sufficiently long
integration times ($>10^{10}$ time unit) the system reach a stationary value.

In Fig.2, we plot the heat flux $J$  as a function of the system size $N$. For
the case of irrational angles, the best fit gives $J\propto N^{-\gamma}$, with
$\gamma = 0.99\pm 0.01$. The  coefficient of thermal conductivity is therefore
independent on $N$, which means that the Fourier law is obeyed.

As a  consistency check  we have  also performed an  independent check  of the
above  results  via  a   Green-Kubo  type  formalism\cite{triang}  from  which
consistent results with those presented above have been obtained.

\begin{figure}[!t]
\includegraphics[width=8.5cm]{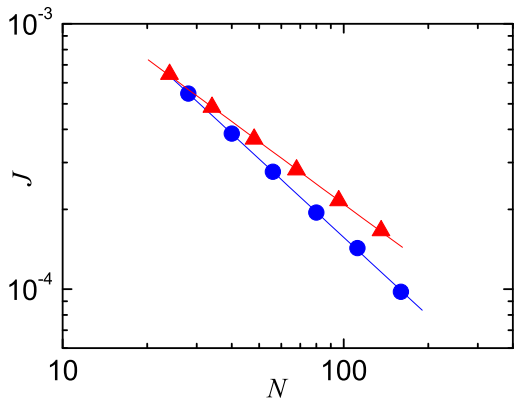}
\caption{
  Scaling behavior of the stationary heat flux $J$ as a function of the system
  size  for the case  of irrational  angles of  Fig. 1  (circles) and  for the
  rational  case (triangles).   $N$ is  the number  of fundamental  cells. The
  particle density was set to $1$ particle per cell independently of $N$.  The
  best-squares fitting gives a slope $ -0.99 \pm 0.01$ for the irrational case
  and $-0.78\pm0.01$ for the rational one.}
\label{heatflux}
\end{figure}

A  completely different  behavior is  obtained when  the angles  $\theta$ and
$\phi$  are rational  multiples of  $\pi$.  The  case with  $\theta=\pi/5$ and
$\phi=\pi/3$,  is  shown  in  Fig.~\ref{heatflux} (triangles),  from  which  a
divergent behavior of  the coefficient of thermal conductivity  $ \kappa \sim
N^{0.22}$ is observed, indicating the absence of Fourier law.

In conclusion,  when all  angles are irrational  multiples of $\pi$  the model
shown in  Fig. 1 exhibits  Fourier law of  heat conduction together  with nice
diffusive  properties.  However,  when  all angles  are  rational multiple  of
$\pi$, the  model shows  abnormal diffusion and  the heat conduction  does not
follow the Fourier law.

One may  argue that the  model considered here  is somehow artificial  and far
from realistic  physical models.  Indeed, noninteracting  particle systems are
certainly less realistic  as in general, LTE is  not established.  However the
problem  under discussion  is quite  delicate and  controversial and  our main
purpose  is  to  understand  which  dynamical  properties  are  necessary  and
sufficient to derive Fourier law.

In  this respect  we  mention two  different  billiards like  models that  are
genuinely  interacting  many-particle  models  and  that  share  some  of  the
properties  of  the model  discussed  here.  The  first  one  is the  \emph{1d
  hard-point   particles  with   alternating  masses}   that  consists   of  a
one-dimensional chain  of elastically colliding free  particles with alternate
masses  $m$  and $M$  \cite{hatano99,Dhar01}.   For  this  model all  Lyapunov
exponents are zero like the  irrational triangle channel.  In \cite{caspr} the
problem of heat conduction in this  model was considered. It was found that in
the  alternating mass  model the  heat current  follows a  power-law behavior
$J\propto   N^{-\alpha}$  with  $\alpha\sim0.745$.    In  contrast   with  the
irrational triangle channel, the alternating  mass model does not obey Fourier
law.

Therefore we  have two  models which are  both mixing and  without exponential
instability:  (i)  the triangular  billiard  channel  \cite{triang} (see  also
\cite{triangle}), which exhibits Fourier law  and (ii) the alternate mass hard
point gas  model\cite{caspr} in which the coefficient  of thermal conductivity
diverges with the system size.  The  difference between the two models is that
in case (ii) the total momentum is  conserved while in case (i) it is not.  We
recall  that in several  recent papers\cite{hatano99,Campbell,narayan}  it has
been suggested that total momentum conservation does not allow Fourier law and
this may explain  the lack of Fourier law for  the one dimensional alternating
mass model.

To clarify  the role of total  momentum conservation, in  \cite{caspb} a model
which  is identical to  the alternate  mass hard-point  gas but  without total
momentum conservation was considered.  Numerical results clearly indicate that
this model, contrary to the translationally invariant model, obeys the Fourier
law.

In perspective, these results  demonstrate that diffusive energy transport and
Fourier  law can  take place  in marginally  stable  (non-chaotic) interacting
many-particle  systems.  As  a consequence  exponential  instability (Lyapunov
chaos) is not necessary for the establishment of the Fourier law. Furthermore,
they show  that breaking  the total momentum  conservation is crucial  for the
validity of  Fourier law  while, somehow surprisingly,  a less  important role
seems to be played by the degree of dynamical chaos.

\section{Fourier Law and the Onset of Quantum Chaos}
\label{sec:quantum}

In section~\ref{sec:fourier} we have  shown that strong, exponential unstable,
classical chaos  is not  necessary (actually, strictly  speaking, is  not even
sufficient \cite{leprifpu})  for normal transport. In this  connection, a main
feature   of   quantum  motion   is   the   lack   of  exponential   dynamical
instability\cite{casati}. Thus it is interesting to inquire if, and under what
conditions, Fourier law emerges from the laws of quantum mechanics.

We  consider an  Ising chain  of $L$  spins $1/2$  with coupling  constant $Q$
subject  to  a uniform  magnetic  field  $\vec{h}  = (h_x,0,h_z)$,  with  open
boundaries. The Hamiltonian reads
\begin{equation} \label{eq:H}
{\mathcal H} = -Q\sum_{n=0}^{L-2}\sigma^z_n\sigma^z_{n+1} +
\vec{h}\cdot\sum_{n=0}^{L-1}\vec{\sigma}_n \ ,
\end{equation}
where the operators  $\vec{\sigma}_n = (\sigma^x_n,\sigma^y_n,\sigma^z_n)$ are
the  Pauli matrices  for  the $n$-th  spin,  $n=0,1,\ldots L-1$.   We set  the
coupling  constant $Q=2$.   In this  system, the  only trivial  symmetry  is a
reflection    symmetry,    $\vec{\sigma}_n\rightarrow\vec{\sigma}_{L-1-n}$.    
Moreover the direction of the  magnetic field affects the qualitative behavior
of the  system: If  $h_z=0$, the Hamiltonian  (\ref{eq:H}) corresponds  to the
Ising  chain in  a transversal  magnetic  field. In  this case  the system  is
integrable as (\ref{eq:H}) can be mapped into a model of free fermions through
standard Wigner-Jordan  transformations.  When  $h_z$ is increased  from zero,
the system  is no longer  integrable and  when $h_z$ is  of the same  order of
$h_x$ quantum  chaos sets in  leading to a  very complex structure  of quantum
states  as well  as to  fluctuations in  the spectrum  that  are statistically
described by Random Matrix Theory  (RMT) \cite{rmt}.  The system becomes again
(nearly) integrable when $h_z \gg  h_x$.  Therefore, by choosing the direction
of the external field we can explore different regimes of quantum dynamics.

The  onset of  quantum  chaos is  commonly  studied in  terms  of the  nearest
neighbor level spacing distribution  $P(s)$ that gives the probability density
to find two  adjacent levels at a distance $s$.  For  an integrable system the
distribution   $P(s)$   has   typically   a   Poisson   distribution   $P_{\rm
  P}(s)=\exp\left(-s\right)$.  In  contrast, in the quantum  chaos regime (for
Hamiltonians  obeying  time-reversal  invariance),  $P(s)$  is  given  by  the
Gaussian Orthogonal Ensemble  of random matrices (GOE).  In  this case, $P(s)$
is well-approximated by the Wigner surmise $P_{\rm WD}(s) = (\pi s/2)\exp(-\pi
s^2/4)$, exhibiting the so-called ``level repulsion''.

\begin{figure}[!t]
\begin{center}
\includegraphics[scale=0.32]{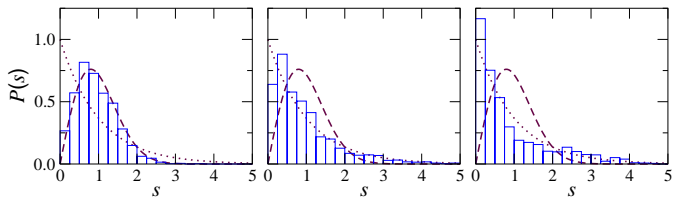}
\caption{\label{fig:1} Nearest neighbor  level spacing distribution $P(s)$ for
  the  chaotic (left), intermediate  (center) and  integrable (right)  chains. 
  $P(s)$  was obtained  by diagonalizing  the Hamiltonian  (\ref{eq:H})  for a
  chain of length $L=12$, averaging over even and odd parity sub-spectra.  The
  curves correspond to  $P_{\rm WD}$ (dashed line) and  to $P_{\rm P}$ (dotted
  line).}
\end{center}
\end{figure}

In Fig.~\ref{fig:1} we show the results of our numerical simulations of system
(\ref{eq:H}) for three different values  of the magnetic field: ({\it i}) {\em
  chaotic case} $\vec{h}=(3.375,0,2)$ at  which the distribution $P(s)$ agrees
with GOE and thus corresponds to  the regime of quantum chaos, ({\it ii}) {\em
  integrable  case} $\vec{h}=(3.375,0,0)$,  at which  $P(s)$ is  close  to the
Poisson    distribution,   and   ({\it    iii})   {\em    intermediate   case}
$\vec{h}=(7.875,0,2)$ at which the  distribution $P(s)$ shows a combination of
(weak) level repulsion and an exponential tail.

In order to study energy transport we need to couple both ends of the chain of
spins to thermal reservoirs at  different temperatures.  In \cite{qfl} we have
devised a simple  way to simulate this coupling, namely the  state of the spin
in  contact  with  the  bath   is  statistically  determined  by  a  Boltzmann
distribution with parameter $T$.  Our model for the reservoirs is analogous to
the stochastic  thermal reservoirs defined by Eqs.~\ref{eq:baths}  and we thus
call it a  \emph{quantum stochastic reservoir}.  In what  follows we use units
in which Planck and Boltzmann constants are set to unity $\hbar=k_{\rm B}=1$.

The  dynamics of the  spins is  obtained from  the unitary  evolution operator
$\mathrm{U}(t)  =   \exp(-i\mathcal{H}t)$.   Additionally  the   leftmost  and
rightmost spins of  the chain are coupled to  quantum stochastic reservoirs at
temperatures $\beta_{\rm l}^{-1}$ and $\beta_{\rm r}^{-1}$ respectively.

The action  of the quantum reservoirs  is as follows: at  times $t=n\tau$ with
$n$  integer the boundary  spins couple  to the  quantum reservoirs.   A local
measurement of the  boundary spins is performed and their new  state is set to
the state  {\em down}  ({\em up}) with  probability $\mu$ ($1-\mu$)  where the
probability $\mu(\beta)$ depends  on the canonical temperature of  each of the
thermal reservoirs as
\begin{equation} \label{eq:canonical}
\mu(\beta_j)  =   \frac{e^{\beta_jh}}{e^{-\beta_jh}  +  e^{\beta_jh}}
\quad ; \quad j \in \{\mathrm{l,r}\} \ .
\end{equation}

The value  of $\tau$  controls the strength  of the  coupling to the  bath. In
\cite{qfl}  it  was   shown  that  averages  over  the   ensemble  of  quantum
trajectories  or time  averages over  a quantum  trajectory are  sufficient to
reconstruct  a thermal density  matrix operator  that correctly  describes the
internal thermal state of the system  in and out of equilibrium. Moreover, the
time averaged expectation values of the  local energy density can be used as a
consistent canonical local temperature.
In order to  compute the energy profile we  write the Hamiltonian (\ref{eq:H})
in terms of local energy density operators $H_n$:
\begin{equation} \label{eq:H_local}
H_{n}     =   -Q\sigma^z_n\sigma^z_{n+1}    +
\frac{\vec{h}}{2} \cdot \left(\vec{\sigma}_n  + \vec{\sigma}_{n+1}\right) \ .
\end{equation}
The local  Hamiltonian $H_n$  (defined for $0  < n  < L-2$), gives  the energy
density   between   the   $n$-th   and   $(n+1)$-th  spins.    In   terms   of
eq.~(\ref{eq:H_local}) the Hamiltonian of the system can be rewritten as
\begin{equation} \label{eq:Htot}
\mathcal{H} = \sum_{n=0}^{L-2}H_n +
\frac{h}{2}(\sigma_\mathrm{l} + \sigma_\mathrm{r}) \ .
\end{equation}

\begin{figure}[!t]
\begin{center}
\includegraphics[scale=0.35]{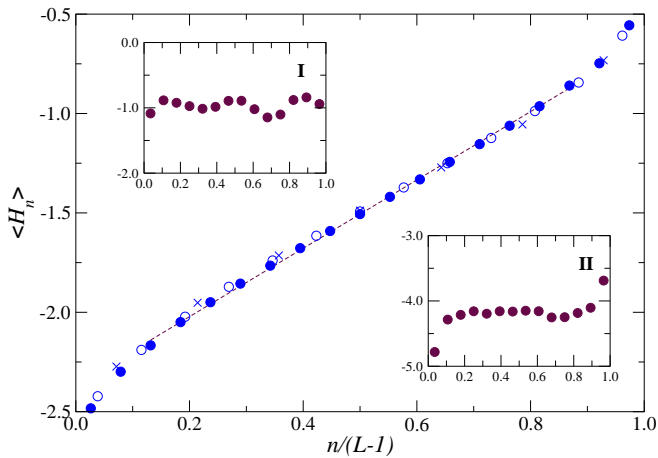}
\caption{
  Out of  equilibrium energy profile  $\ave{H_n}$ for the chaotic  chain.  The
  temperatures of the baths were set to $T_{\rm  l} = 5$ and $T_{\rm r} = 50$.
  Results for chains of size $L=8$ (crosses), $L=14$ (open circles) and $L=20$
  (solid circles) are  shown.  The dashed line was obtained  from a linear fit
  of the data for $L=20$ for the $L-4$ central spins. Insets (I) and (II) show
  the energy  profile for the integrable and  intermediate cases respectively,
  for $L=15$.
\label{fig:3}}
\end{center}
\end{figure}

In  Fig.~\ref{fig:3} we  show the  energy profile  $\ave{H_n}$ for  an  out of
equilibrium  simulation   of  the  chaotic  chain.    In  all  non-equilibrium
simulations,  the temperatures  of the  baths were  set to  $T_{\rm  l}=5$ and
$T_{\rm r}=50$.  After an appropriate scaling the profiles for different sizes
collapse to  the same curve.  More interesting,  in the bulk of  the chain the
energy profile  is in  very good approximation  linear.  In contrast,  we show
that  in the  case of  the integrable  (inset I)  and intermediate  (inset II)
chains, no energy gradient is created.

We now define the local  current operators through the equation of continuity:
$\partial_t{H_n} =  i[{\mathcal H},H_n] =  - (J_{n+1} - J_n)$,  requiring that
$J_n = [H_n,H_{n-1}]$.  Using eqs.  (\ref{eq:H_local}) and (\ref{eq:Htot}) the
local current operators are explicitly given by
\begin{equation}
J_{n} = h_xQ\left(\sigma_{n-1}^z-\sigma_{n+1}^z\right)\sigma^y_{n},
\quad
1\le n\le  L-2.
\end{equation}

\begin{figure}[!t]
\begin{center}
\includegraphics[scale=0.35]{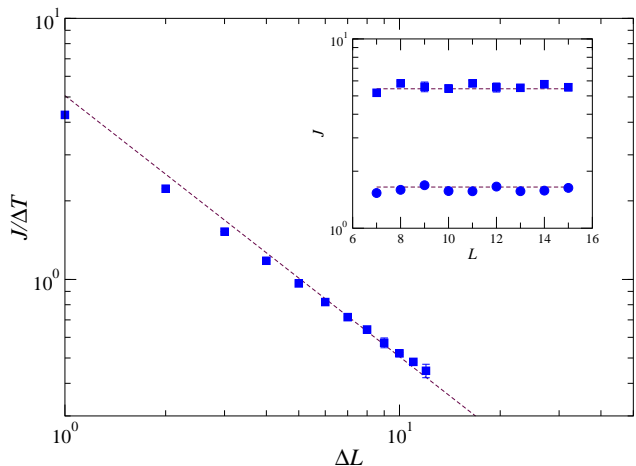}
\caption{\label{fig:4}
  Size dependence of the energy current in the chaotic chain with $T_{\rm l} =
  5$  and $T_{\rm  r}  = 50$.  The  dashed line  corresponds  to $1/\Delta  L$
  scaling.  In the  inset, the size dependence of the  energy current is shown
  for the integrable (circles) and the intermediate (squares).}
\end{center}
\end{figure}

In Fig.~\ref{fig:4} we plot $J/\Delta T$ as  a function of the size $L$ of the
system  for sizes up  to $L=20$.   The mean  current $J$  is calculated  as an
average of $\ave{J_n}$  over time and the $L-8$ central  values of $n$.  Since
$\Delta  L=L-8$ is an  effective size  of the  truncated system,  the observed
$1/\Delta L$ dependence  confirms that the transport is  normal.  On the other
hand, in integrable and intermediate  chains we have observed that the average
heat current does not depend on  the size $J\propto L^0$, clearly indicating a
ballistic transport.

In conclusion we have shown that Fourier law of heat conduction can be derived
from the  pure quantum dynamical evolution without  any additional statistical
assumption. Our  results suggest that onset  of quantum chaos  is required for
the validity of Fourier law.

\section{Thermal Rectification}
\label{sec:rectifier}

We now turn our attention to the possibility to control heat flow. To this end
we start with the model introduced in \cite{diod} and consider the Hamiltonian
\begin{equation}
H = \sum_{n=1,N} \frac{p_n^2}{2 m} + V_n(y_n) + \frac{1}{2} K
(y_n-y_{n-1})^2
\label{PBmodel}
\end{equation}
\noindent
which describes  a chain of $N$  particles with harmonic  coupling of constant
$K$ and a Morse on-site potential $V_n(y_n)=D_n(e^{-\alpha_n y_n}-1)^2$.  This
model was introduced  for DNA chains where  $m$ is the reduced mass  of a base
pair, $y_n$ denotes  the stretching from equilibrium position  of the hydrogen
bonds connecting  the two bases of the  $n$-th pair and $p_n$  is its momentum
\cite{peyrard}.  In  the context of  the present study,  model (\ref{PBmodel})
can simply  be viewed as a  generic system of  anharmonic coupled oscillators,
the  on-site potential  arising  from  interactions with  other  parts of  the
system, not included  in the model.  The Morse potential  is simply an example
of a  highly anharmonic soft potential,  which has a  frequency that decreases
drastically when the amplitude of the motion increases.

We  consider the  out-of-equilibrium  properties of  model (\ref{PBmodel})  by
numerically simulating the dynamics of  the $N$ particle chain, coupled at the
two  ends, with  thermal baths  at different  temperatures.  We  thermalize at
$T_L$ and  $T_R$ the first and  the last $L$ particles  by using Nos\'e-Hoover
thermostats chains \cite{Nose}, or  a Langevin description when we investigate
cases  very far  from equilibrium.   The baths  temperatures $T_L$,  $T_R$ are
never  large  enough to  drive  the  system  beyond the  thermal  denaturation
temperature $T_c$ above which the mean value of $y_n$ diverges \cite{pey3}.

\begin{figure}[!t]
\includegraphics[width=8cm]{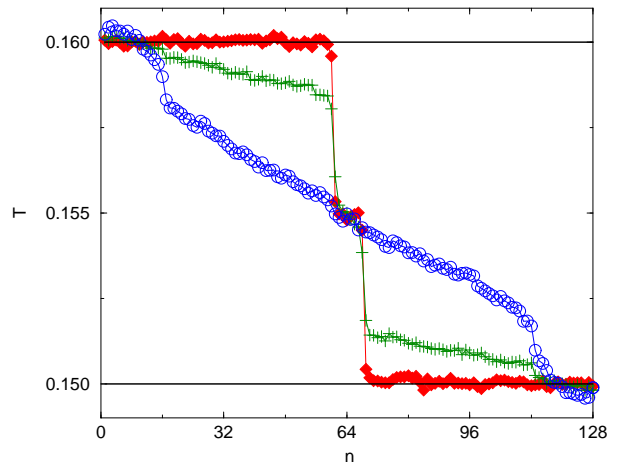}
\caption{\label{fig:distance}
  Temperature  profile of  the chain  with parameters  $D_L=D_R=0.5$, $N=128$,
  $\alpha=1$, $K=0.3$, $M=8$, and  $T_L=0.16$, $T_R=0.15$. The Morse potential
  constants   are  $D_0=0.5$  -   homogeneous  system-   (circles),  $D_0=1.2$
  (diamonds) and $D_0=0.8$ (+)}
\end{figure}

We  compute  the  temperature  profile  inside the  system,  i.e.   the  local
temperature at  site $n$ defined  as $T_n =  m \langle \dot{y_n}^2  \rangle $,
where $\langle  \hspace{0.2cm} \rangle$ stands  for temporal average,  and the
local  heat flux  $J_n =  K \langle  \dot{y_n} (y_n  -  y_{n-1})\rangle$.  The
simulations  are  performed  long  enough  to  allow the  system  to  reach  a
non-equilibrium steady state  where the local heat flux  is constant along the
chain.

As a preliminary step we have  considered the homogeneous case in which $D_n =
D$,  $\alpha_n=\alpha$,  $n=1,..N$. Here,  as  expected,  we  have detected  a
temperature gradient  inside the chain, and  we have verified  that the system
obeys the Fourier law of heat conduction \cite{diod}.

For  the heterogeneous case  we divide  the chain  between the  thermostats in
three regions  in which $D_n$ takes  different values.  In the  left and right
regions, $D_n=D_L$ for $n=1,\ldots (N  - M)/2$ and $D_n=D_R$ for $n=(N+M)/2+1,
\ldots, N$,  respectively.  For  the $M$  sites of the  central region  we set
$D_n=D_0$. For the whole chain $\alpha_n = \alpha$.

In Fig.~\ref{fig:distance}  \cite{diod}, we show the  temperature profiles for
three different  values of  $D_0$.  Clearly a  conductor--insulator transition
occurs in the heterogeneous system as a function of the inhomogeneity measured
by $D_0$. We  have obtained that the averaged heat flux  $J$, which is uniform
along  the  whole chain  in  the  steady state,  decreases  by  two orders  of
magnitude  when  $D_0-D$ ($D=D_L=D_R$)  increases  from  $0$  to $0.7$.   This
transition is  explained in terms  of a linearized version  of (\ref{PBmodel})
\cite{diod}: for the  nonlinear Hamiltonian Eq.~\ref{PBmodel}, {\em effective}
phonon bands can be defined whose  size and position depend on the microscopic
properties of  the interaction (in particular  of coefficient $D$)  and on the
temperature.   Using this  dependence in  a smart  way one  can design  a {\em
  thermal rectifier}  by choosing  appropriate parameters $D_L$,  $D_R$, $D_0$
and $\alpha$,  depending on the range  of temperatures in  question. We intend
for a  thermal rectifier a device  in which, for a  fixed external temperature
gradient,  the magnitude  of  the heat  current  depends on  the  sign of  the
gradient.

\begin{figure}
\includegraphics[bb=0 -30 654 462, width=\columnwidth]{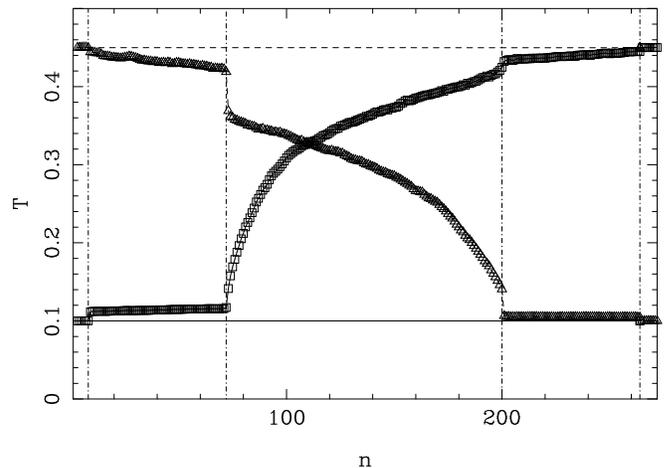}
\vspace{-1cm}
\caption{
  Temperature   profiles  in   a  ``thermal   rectifier''  for   two  opposite
  orientations of the thermal gradient. The dash-dotted lines show the borders
  of the different regions in the  lattice. The thermostats have a size $L=8$,
  the central region a size $M=128$, and the left and right regions contain 64
  particles.  The  coupling constant  is $K=0.18$, and  the parameters  of the
  Morse potential are $D_L=4.5$,  $\alpha_L=0.5$, $D_0=0.7$, $\alpha_0 = 1.4$,
  $D_R  = 2.8$,  $\alpha_R  = 0.5$  in  the left,  central  and right  regions
  respectively.  The temperatures  of the thermostats are $T=0.1$  and $0.45$. 
  When the high  temperature is on the right side of  the lattice, the average
  flux is $J = 0.146~10^{-3}$ while, when the thermal gradient is reversed the
  flux drops to  $J=0.755~10^{-4}$.  Note in this case  the discontinuities at
  the interfaces  between region  which attest of  the bad energy  transfer at
  these points.
\label{fig:diode}}
\end{figure}

For instance, for  the heterogeneous model one can  design a thermal rectifier
by  setting  a  strongly   nonlinear  region  sandwiched  between  two  weakly
anharmonic left  and right domains. In  the presence of a  thermal gradient in
the central  part, the effective phonon  frequencies evolve in space  in a way
that depends  on the orientation of  the gradient.  This can  provide either a
good matching of the bands at  the interfaces, leading to a thermal conduction
across  the  system,  or  a  complete  mismatch  leading  to  poor  conduction
\cite{diod}.  We  exemplify this  situation in Fig.~\ref{fig:diode}  where the
temperature  profiles for  the two  orientations of  the thermal  gradient are
shown.   For the parameters  values of  Fig.~\ref{fig:diode} the  heat current
changes by a factor  of about 2 when the direction of  the thermal gradient is
reversed.

Recently  we  have shown  that  using  different  microscopic models  for  the
interaction  like {\em  e.g.}  a  Frenkel-Kontorova on-site  potential,  it is
possible  to strongly  improve the  efficiency of  the rectifier  \cite{sing}. 
Even though  the underlying microscopic mechanism discussed  in \cite{sing} is
different from  the one in \cite{diod} it  is based on the  same general idea:
the temperature dependence  of the phonon band. This dependence  is due to the
nonlinearity of the  potential and therefore it should  be possible to observe
the rectifying effect in any nonlinear lattice.

But thermal rectification is not exclusive to nonlinear lattices.  Recently we
also have shown that this phenomenon  can also be observed in billiard systems
of interacting  particles \cite{EMM}.   This model is  based on  the effective
interaction of  a gas of  particles with the  scatterers of a  Lorentz channel
\cite{mejia-1,mejia-2}.  We  have shown that the interacting  character of the
particles  gives rise to  dynamical memory  effects that  depend on  the local
thermodynamical  fields, in  particular on  the local  temperature  gradients. 
These memory effects have been  described in detail in \cite{emmz,EY}. We have
used this dependence  to set up a particular geometry  for which large thermal
rectifications  (the  heat  currents  for  both orientations  of  the  thermal
gradient differ by  several orders of magnitude) are  obtained \cite{EMM}. The
possibility to obtain large rectification of the heat flow in billiard systems
has  raised  a  great  interest   because  they  are  more  easily  realizable
experimentally in the rapidly expanding field of nanophysics.

\section{Conclusions}
\label{sec:concl}

We have discussed the problem of  heat conduction in classical and quantum low
dimensional systems in relation to  the dynamical properties of the system. At
the classical  level we  have provided convincing  numerical evidence  for the
validity  of Fourier law  of heat  conduction in  linear mixing  systems, {\em
  i.e.}   in  systems  without  exponential  instability.   As  a  consequence
exponential   instability  (Lyapunov   chaos)   is  not   necessary  for   the
establishment of  the Fourier law. Moreover,  we have shown  that breaking the
total momentum conservation is crucial  for the validity of Fourier law while,
somehow surprisingly, a  less important role seems to be  played by the degree
of dynamical chaos.

At  the quantum  level,  where the  motion  is characterized  by  the lack  of
exponential  dynamical instability,  we have  shown that  Fourier law  of heat
conduction can  be derived from  the pure quantum dynamical  evolution without
any  additional  statistical assumption.   Similarly  to  our observations  in
classical models, our results for a chain of interacting spins suggest that in
quantum mechanics, which is characterized by the lack of exponential dynamical
instability,  the onset  of  quantum chaos  is  required for  the validity  of
Fourier law.

We have  also discussed the  phenomenon of thermal rectification  in different
classical  models  and  discussed  different types  of  microscopic  classical
mechanisms that lead to the rectification of heat flow.  Even though all these
mechanism  are  rooted in  one  way  or another  in  the  nonlinearity of  the
microscopic dynamics, we believe that, as for the case of Fourier law, thermal
rectification  is also  possible  at the  quantum  level. Some  steps in  this
direction have already be taken \cite{saito,segal,pmmc}

\begin{acknowledgments}
  We  gratefully   acknowledge  support   by  the  MIUR-PRIN   2005  ``Quantum
  computation with  trapped particle arrays,  neutral and charged''. C.  M.-M. 
  acknowledges  a  Lagrange  fellowship  from  the  Institute  for  Scientific
  Interchange Foundation.
\end{acknowledgments}

\end{document}